\documentclass[journal=ancac3,manuscript=article]{achemso}

\usepackage[utf8]{inputenc}
\usepackage[T1]{fontenc}       
\usepackage{mathtools}
\usepackage{amsmath}
\usepackage{amsfonts}
\usepackage{chemformula}
\usepackage{hyperref}

\author{Stefano~{Dal~Forno}}
\email{sdalforno@ntu.edu.sg}
\affiliation{School of Physical and Mathematical Sciences, Nanyang Technological University, 21 Nanyang Link, Singapore 637371}

\author{Natsumi~Komatsu}
\affiliation{Department of Electrical and Computer Engineering, Rice University, Houston, TX 77005, USA}

\author{Michael~Wais}
\affiliation{School of Physical and Mathematical Sciences, Nanyang Technological University, 21 Nanyang Link, Singapore 637371}
\alsoaffiliation{Institute of Solid State Physics, TU Wien, Vienna, Austria}

\author{Ali~Mojibpour}
\affiliation{Department of Electrical and Computer Engineering, Rice University, Houston, TX 77005, USA}

\author{Indrajit~Wadgaonkar}
\affiliation{School of Physical and Mathematical Sciences, Nanyang Technological University, 21 Nanyang Link, Singapore 637371}

\author{Saunab~Ghosh}
\affiliation{Department of Electrical and Computer Engineering, Rice University, Houston, TX 77005, USA}

\author{Yohei~Yomogida}
\author{Kazuhiro~Yanagi}
\affiliation{Department of Physics, Tokyo Metropolitan University, Hachioji, Tokyo 192-0397, Japan}

\author{Karsten~Held}
\affiliation{Institute of Solid State Physics, TU Wien, Vienna, Austria}

\author{Junichiro~Kono}
\affiliation{Department of Electrical and Computer Engineering, Rice University, Houston, TX 77005, USA}
\alsoaffiliation{Department of Physics and Astronomy, Rice University, Houston, TX 77005, USA}
\alsoaffiliation{Department of Materials Science and NanoEngineering, Rice University, Houston, TX 77005, USA}
\alsoaffiliation{Smalley-Curl Institute, Rice University, Houston, TX 77005, USA}
\alsoaffiliation{School of Physical and Mathematical Sciences, Nanyang Technological University, 21 Nanyang Link, Singapore 637371}

\author{Marco~Battiato}
\email{marco.battiato@ntu.edu.sg}
\affiliation{School of Physical and Mathematical Sciences, Nanyang Technological University, 21 Nanyang Link, Singapore 637371}

\title[CNT background absorption from Boltzmann equation]{Origin of the Background Absorption in Carbon Nanotubes: Phonon-Assisted Excitonic Continuum}


\begin{document}

\begin{tocentry}
\includegraphics{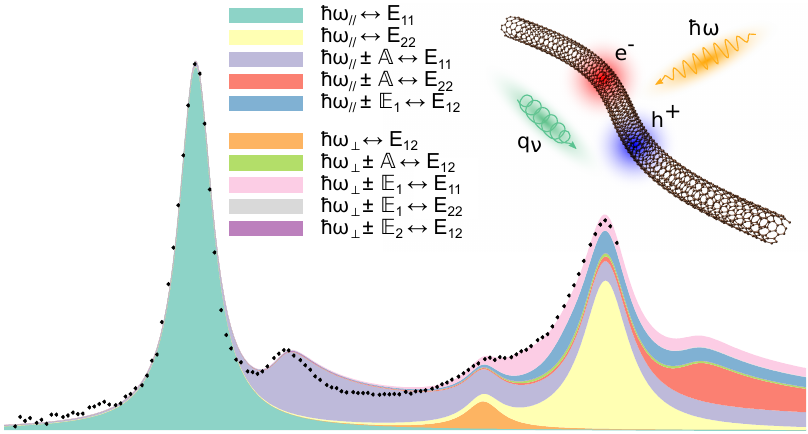}
\end{tocentry}

\newpage
\begin{abstract}

Excitonic effects in 1D semiconductors can be qualitatively different from those in higher dimensions. 
In particular, the Sommerfeld factor, the ratio of the above-band-edge excitonic continuum absorption to free electron-hole pair generation, has been shown to be less than 1 (i.e., suppressed) in 1D systems while it is larger than 1 (i.e., enhanced) in 2D and 3D systems.
Strong continuum suppression indeed exists in semiconducting single-wall carbon nanotubes, a prototypical 1D semiconductor. However, absorption spectra for carbon nanotubes are typically fit with a combination of Lorentzians and a polynomial background baseline with little physical meaning.
Here, we performed absorption measurements in aligned single-chirality (6,5) carbon nanotube films. The obtained spectra were fit with our theoretical model obtained by solving the Boltzmann scattering equation (i.e., the quantum Fokker-Planck equation), involving fifty-nine different types of transitions among three different types of quasiparticles.
Specifically, we took into account microscopic interactions between photons, phonons, and excitons, including their dispersions, which unambiguously demonstrated that the background absorption is due to phonon-assisted transitions from the semiconductor vacuum to finite-momentum continuum states of excitons.
The excellent agreement we obtained between experiment and theory suggests that our numerical technique can be seamlessly extended to compute strongly out-of-equilibrium many-body dynamics and time-resolved spectra in low-dimensional materials.

\end{abstract}

\newpage
\section{Introduction}

Excitons in one-dimensional (1D) systems are expected to behave significantly differently from those in other dimensions.  For example, the exciton binding energy becomes infinite in an ideal 1D electron-hole system~\cite{loudon_one-dimensional_1959,elliott_theory_1959,elliott_theory_1960}.  In addition, 1D Mahan excitons are expected to be robustly insensitive to the dynamics of holes, whereas the counterparts in higher dimensions are known to be readily destroyed by hole recoil.\cite{ogawa_fermi-edge_1992} Further, the Sommerfeld factor has been shown to be less than 1 in 1D systems,\cite{ogawa_interband_1991, ogawa_optical_1991} suggesting suppression of the excitonic continuum absorption above the band edge compared to the strength of free electron-hole pair generation. This is in stark contrast to 2D and 3D excitons, for which the Sommerfelt factor is larger than 1, corresponding to excitonic enhancement of absorption.\cite{haug_quantum_2009}

Single-wall carbon nanotubes (SWCNTs)~\cite{jorio_carbon_2008} provide model 1D systems in which to address fundamental questions in condensed matter physics. 
Recent advances in separation, sorting, and assembling techniques have allowed researchers to prepare ordered macroscopic ensembles of single-chirality SWCNTs.\cite{gao_science_nodate} 
Optical spectroscopic and optoelectronic properties of semiconducting SWCNTs\cite{weisman_handbook_2011,nanot_optoelectronic_2012} are naturally affected by strong 1D excitonic effects~\cite{ando_excitons_1997, chang_excitons_2004, spataru_excitonic_2004, perebeinos_scaling_2004, kane_electron_2004}
Asymmetric peaks, expected from 1D van Hove singularities, as well as continuum absorption above the band-edge, are strongly suppressed, leaving symmetric excitonic absorption peaks. Two-photon photoluminescence excitation measurements~\cite{maultzsch_exciton_2005, wang_optical_2005} have shown exciton binding energies to be in the $\sim$300~meV range for $\sim$1~nm diameter SWCNTs.  

Despite much experimental progress made during the past decade in characterising the optical and excitonic properties of SWCNTs, modelling such properties is an extremely challenging task. It requires a description of the nontrivial interplay between photons, electrons, holes, excitons, and phonons. Theoretical approaches involving accurate many-body techniques face the issue of the prohibitive scaling of the numerical cost and complexity, and often have to rely on close-to-equilibrium approximations. Further, they generally focus on a single type of scatterings by disentangling the dynamics of several subsystems \cite{capaz_diameter_2006, spataru_theory_2005, torrens_energy_2008, jiang_chirality_2007}.
For these reasons, experimental optical spectra are commonly fit using a combination of Lorentzians and polynomials \cite{pfohl_fitting_2017, katsutani_direct_2019}. However, the baseline background polynomial has no physical meaning, and fitting results depend on the choice of functions. This procedure therefore not only prevents the understanding of the origin of the background but also degrades the quality of the information extracted from the main peaks.

Potentially, even continuous wave (CW) linear optical absorption measurements can provide a wealth of information about dynamics and interactions.
For example, the relative amplitude between the main excitonic peaks and the phonon-assisted peaks\cite{perebeinos_effect_2005} in SWCNT absorption spectra could be used for studying the time evolution of phonon populations \cite{yu_phonon_2010}.
Yet, even more information can be harvested: the relaxed momentum conservation condition due to the presence of phonons allows for transitions to excitonic states away from the band bottom. Such transitions give an insight into the population of higher-energy excitonic states as well as phononic ones. However, such contributions are usually encrypted in the tails of the main peaks\cite{torrens_energy_2008}.
It is therefore critical to have a theoretical framework that is capable of extracting this information and providing a consistent description of experimental results.

In this work, we measured optical absorption in films of aligned (6,5) SWCNTs and then fitted the results by solving the quantum Boltzmann equation.
We included the coupling of photons, phonons, and excitons in SWCNTs and solved the Boltzmann equation explicitly. Our approach is capable of modelling strongly out-of-equilibrium populations and high-order scatterings, by leveraging a newly developed numerical algorithm that dramatically reduces the computational cost of the scattering integrals. Results showed excellent agreement with experiments and showed the capabilities of our numerical approach to transport and scattering problems.
Importantly, we identified the origin of the background baseline absorption to be phonon-assisted continuum absorption through transitions to finite-momentum exciton states. Furthermore, our method is ready to be extended, without any further implementation, to  calculating  both the time evolution of particle distributions of complex systems involving several heterogeneous scattering types and without any close-to-equilibrium approximation and time resolved spectra \cite{wais_numerical_2021,wais_quantum_2018,bagsican_terahertz_2020,wais_comparing_2021}.


\section{Results and discussion}

\textbf{Description of the model.}
A system of interacting (quasi)particles can be modelled in terms of the Boltzmann equation (BE). Each particle is described by a population function $f_n(\mathbf{k})$ where $n$ is an index that includes both particle type and band number and $\mathbf{k}$ is the reciprocal lattice vector. In the absence of spatial resolution and external fields, the BE reduces to
\begin{equation}
    \frac{ \partial f_n}{ \partial t} = \sum_{\alpha} {\left( \frac{ \partial f_n}{ \partial t} \right)}_{\alpha}
\end{equation}
where the scattering integrals on the right hand side are written in terms of the quantum Fokker–Plank equation \cite{snoke_solid_2020} (see Methods).

To account for the coupling with light, we model the (6,5)~SWCNTs as a system of weakly interacting excitons, phonons, and photons. The quantum Fokker-Plank equation is derived assuming that the quasiparticles have either fermionic or bosonic commutation relations. Excitons behave as bosons or fermions in the limit of low or high occupation, respectively. In the present work, we will limit ourselves to the regime of fermionic excitons \cite{hanamura_condensation_1977} (yet, this choice has a negligible impact on the results).

\begin{figure}[tb]
    \centering
    \includegraphics[]{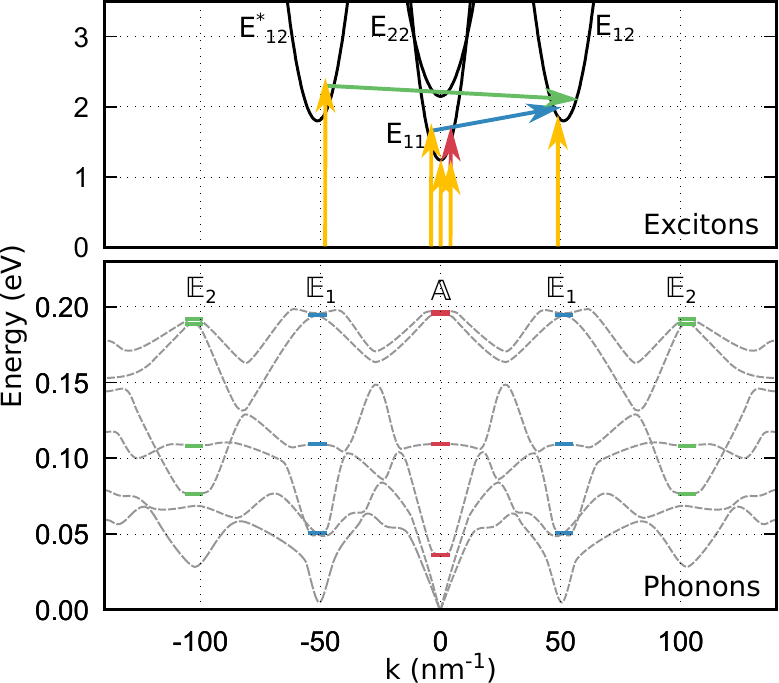}
    \caption{Excitonic and phononic bands in the extended helical coordinates. We remind that effective masses for the excitons have been magnified 50 times for clarity. Phonon energies are shown by the solid color horizontal lines. The dashed grey lines are purely guide to the eye since no data is available in that region. Color arrows on the top figure show some of the possible excitation paths. Yellow vertical arrows represent photon absorption, while red, blue and green diagonal arrows show some possible scattering due to $\mathbb{A}$, $\mathbb{E}_1$ and $\mathbb{E}_2$ phonons, respectively.}
    \label{fig:exc_phn_bands}
\end{figure}

Figure~\ref{fig:exc_phn_bands} shows the exciton, phonon, and photon band structure used in this study plotted in helical coordinates.
The helical representation dramatically simplifies the description by reducing the number of subbands and allowing for an easy accounting of optical selection rules.
A thorough description of the helical and linear coordinate representations along with a detailed review of symmetry properties of SWCNTs can be found in the work of Barros \textit{et al}.~\cite{barros_review_2006,barros_selection_2006}, while a brief summary is given in the Supporting Information for the reader's convenience.

The photon energies have a dispersion relation of the form $\epsilon(k)$, where $k$ is the projection of the photon wavevector on the nanotube axis. For optical frequencies the momentum transferred by the photon is negligible as the transition is practically vertical.
Photons polarised parallel to the SWCNT's axis (pht$_{//}$) have the usual photon dispersion $\epsilon_{//}(k) = c \hbar  | k |$.
On the other hand, the dispersion of perpendicularly polarised photons (pht$_{\perp}$) is $\epsilon_{\perp}(k) = c \hbar | k - \Delta k_{\perp} | $ where $\Delta k_{\perp} = 50.99$~nm$^{-1}$ is the pseudo-momentum carried by the perpendicular polarised photon arising from the helical representation.

We include in our description of the (6,5)~SWCNTs four low energy spin singlet bright excitonic bands, following Ref.~\cite{dresselhaus_exciton_2007}. We use a parabolic approximation close to the bottom of the band to resemble the \textit{ab initio} results obtained with the Bethe--Salpeter equation (BSE). The exciton effective masses have been approximated by the sum of the electron and hole effective masses in (6,5) SWCNTs~\cite{mattis_what_1984}. In particular, this approximation holds for Wannier--Mott excitons such as those in (6,5) SWCNTs~\cite{luer_size_2009}. We include the singlet excitons $E_{11}$ and $E_{22}$ originated from electrons and holes both in the K (or K') valley. In the top panel of Fig.~\ref{fig:exc_phn_bands}, they appear as parabolas centred around $k=0$ with effective masses $\mu_{11} = 0.21$ and $\mu_{22} = 0.44$ in units of the bare electron mass $m_e$, and energy minima $E^{min}_{11}=1.23$ and $E^{min}_{22}=2.15$~eV, respectively~\cite{dresselhaus_exciton_2007}. Note that, due to the size of the helical zone, the dispersion of all excitonic bands appears extremely narrow: we therefore expanded them horizontally by a factor of 50, for easier visualisation. Since the total pseudo-momentum must be conserved, $E_{11}$ and $E_{22}$ excitons can be excited only by photons parallel to the SWCNT's axis, but not by the perpendicular ones.

The other two exitonic bands are the cross polarisation singlet $E_{12}$ and $E^*_{12}$. We use $\mu_{12} = 0.32$ and $E^{min}_{12}=1.86$ eV~\cite{dresselhaus_exciton_2007}. $E_{12}$ and $E^*_{12}$ are formed by a linear combination of electrons and holes coming from the same valley but separated by $\Delta k_{\perp}$ in the helical BZ (Fig.~S1 in the Supporting Information). Because of chirality, in the specific case of (6,5) SWCNTs, the magnitude of $\Delta k_{\perp}$ is smaller than the reciprocal space distance between two adjacent K-point valley $\Delta k_{KK}$. This means that the bright cross-polarised excitons $E_{12}$ cannot be excited at the bottom of the band. Only photons with polarisation perpendicular to the CNT's axis can excite $E_{12}$ or $E^*_{12}$ excitons. Dark excitons are not considered.

Several phonon modes are included in the calculations. We explicitly include only the region of the helical BZ that can contribute to phonon-assisted optical generation of excitons.\cite{perebeinos_effect_2005} These regions are highlighted by coloured bars in Fig.~\ref{fig:exc_phn_bands}. The dashed lines are only a guide for the eye, to remind the reader that the selected phonon energies are part of a dispersion extending across the helical BZ. Phonon energies have been taken from the work of Lim \textit{et al}.~\cite{lim_ultrafast_2014}. Specifically, we included only the first-order Raman-active phonons and labelled them using the notation of the irreducible group of the lattice modes\cite{barros_review_2006}, namely $\mathbb{A}^i$, $\mathbb{E}_1^i$ and $\mathbb{E}_2^i$ symmetry phonons, where $i=3,4,5,6$ labels the $i$-th phonon mode. Furthermore, we neglect contribution to the absorption coming from the two lowest acoustic phonons due to their vanishing optical matrix elements~\cite{lim_ultrafast_2014}.
All phonon modes with their standard nomenclature and energies in cm$^{-1}$ are summarised in Table~\ref{tab:phonon}.
\begin{table}
\centering
\begin{tabular}{c|ccc}
mode no. & $\mathbb{A}$ & $\mathbb{E}_1$ & $\mathbb{E}_2$ \\ 
\hline
3rd & 294 (RBM) & 407 & 616 \\ 
4th & 884 (oTO) & 881 & 874 \\
5th & 1575 (G--) & 1568 & 1521 \\
6th & 1588 (G+) & 1570 & 1548
\end{tabular}
\caption{Phonon modes and energies in cm$^{-1}$.}
\label{tab:phonon}
\end{table}

\begin{figure} [tb]
    \centering
    \includegraphics[]{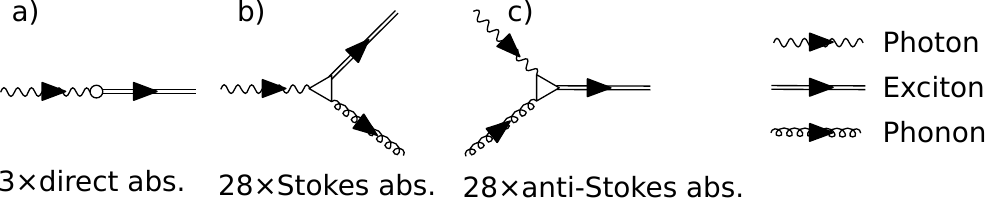}
    \caption{Schematic representation of the scattering processes in Eqs.~\ref{eq:parallel_scatt} and \ref{eq:perpendicular_scatt}.}
    \label{fig:schematic-scatterings}
\end{figure}

\textbf{Scattering channels.} Following the coupling rules predicted by group theory and conservation of pseudo-momentum, we can group the possible absorption scattering channels into 7 categories.
For light polarised parallel to the CNT axis we have
\begin{equation} \label{eq:parallel_scatt}
\begin{split}
i)  \quad & \text{pht}_{//} \longleftrightarrow E_{ii} \\
ii) \quad & \text{pht}_{//} \pm \mathbb{A} \longleftrightarrow E_{ii} \\
iii)\quad & \text{pht}_{//} \pm \mathbb{E}_1 \longleftrightarrow E_{ij} \\
\end{split}
\end{equation}
where $i=1,2$ and $j=1,2$. The first channel refers to direct photon absorption or emission. Channels ii) and iii) are phonon-assisted photon absorption or emission, where the $-$ ($+$) sign correspond to Stokes (anti-Stokes) transitions. 
On the other hand, the scattering families for perpendicularly polarised light are
\begin{equation} \label{eq:perpendicular_scatt}
\begin{split}
iv) \quad & \text{pht}_{\perp} \longleftrightarrow E_{ij} \\
v)  \quad & \text{pht}_{\perp} \pm \mathbb{A} \longleftrightarrow E_{ij} \\
vi) \quad & \text{pht}_{\perp} \pm \mathbb{E}_1 \longleftrightarrow E_{ii} \\
vii)\quad & \text{pht}_{\perp} \pm \mathbb{E}_2 \longleftrightarrow E_{ij} \\
\end{split}
\end{equation}
Including all possible combinations of phonon modes in Eqs.~\ref{eq:parallel_scatt} and \ref{eq:perpendicular_scatt}, we can construct a total of 59 different scattering channels: 3 direct $E_{11}$, $E_{22}$ and $E_{12}$ absorption plus 7 $\times$ 4 $\times$ 2 phonon assisted transitions of 4 phonon modes and 2 Stokes and anti-Stokes processes.
These are represented schematically in Figure~\ref{fig:schematic-scatterings}.
Higher-order processes such as double resonance Raman scattering are neglected.
Also, we were unable to find energy dispersion and lifetimes for the $dE_{11}$ dark exciton (or K-momentum dark exciton) and its coupling to K-point phonons (also known as D phonons or zone-boundary phonons). For this reason, the contribution of this process to the absorption was neglected, in spite of the fact that this exciton has been shown to contribute to the photoluminescence and optical absorption of (6,5) SWCNT due to its relatively large effective mass \cite{torrens_energy_2008}.

\textbf{Light absorption and quasi-particle lifetimes.}
The probability rate of absorbing a photon of energy $\hbar \omega = c\hbar k_1$ for a generic scattering channel $\alpha$ is
\begin{equation}
    R_{\alpha}(\mathbf{k}_1) = \frac{\delta}{\delta f_1} {\left( \frac{ \partial f_1}{ \partial t} \right)_{\alpha}}
    \label{eq:scatt-rate}
\end{equation}
where $f_1$ is the population of photons. A detailed derivation and the full expression of Eq.~\ref{eq:scatt-rate} can be found in Ref.\cite{wais_numerical_2021} and in the Methods section. The absorption spectrum is proportional to the sum of all the scattering channels. We define the total parallel and the total perpendicular absorption spectrum $\sigma_{//}(\omega)$ and $\sigma_{\perp}(\omega)$ as the sum of the scattering rates in Eq.~\ref{eq:parallel_scatt} and Eq.~\ref{eq:perpendicular_scatt}, respectively.
Notice that Eq.~\ref{eq:scatt-rate} holds in general and not only close to equilibrium. Therefore it can be used both for static and dynamic spectra by simply substituting the instantaneous populations of the particles involved in the transition. 

For the general case of a quasi 2D sample of anisotropic SWCNTs we define the total absorption as
\begin{equation}
\begin{split}
    \label{eq:total_abs}
    \sigma^{\beta}(\omega) &= \sigma_{//}(\omega) \int_{-\pi/2}^{\pi/2} p(\theta) \cos^2(\theta-\beta) d\theta \; + \\
    & + \sigma_{\perp}(\omega) \int_{-\pi/2}^{\pi/2} p(\theta) \sin^2(\theta-\beta) d\theta,
\end{split}
\end{equation}
where $\beta$ is the polarisation angle of light with respect to the SWCNTs alignment, $p(\theta)$ is the 2D angular distribution density of the SWCNTs and $\theta$ is the angle between the macroscopic alignment of the sample and the individual tubes.
For a given angular distribution, we can rewrite Eq.~\ref{eq:total_abs} for the two interesting cases of $\beta=0$ (parallel sample) and $\beta=\pi/2$ (perpendicular sample)
\begin{equation}
    \label{eq:total_para}
    \sigma^{0}(\omega) = \sigma_{//}(\omega) \, \alpha + \sigma_{\perp}(\omega) \, (1-\alpha),
\end{equation}
\begin{equation}
    \label{eq:total_perp}
    \sigma^{\pi/2}(\omega) = \sigma_{//}(\omega) \, (1-\alpha) + \sigma_{\perp}(\omega) \, \alpha,
\end{equation}
with $\alpha$ given by
\begin{equation}
    \label{eq:alpha}
    \alpha = \int_{-\pi/2}^{\pi/2} p(\theta) \cos^2(\theta) d\theta.
\end{equation}
The mixing parameter $\alpha$ equals 1 for perfectly aligned SWCNTs and equals 1/2 for randomly arranged SWCNTs. We can also define the two-dimensional nematic order parameter $S=2 \alpha -1$ for the orientatrion of the SWCNTs in a similar fashion to the work of Katsutani \textit{et al}.~\cite{katsutani_direct_2019}. Since the angular distribution function of our SWCNT sample is not known a priori, $\alpha$ will be used as a fitting parameter.

The individual absorption spectra obtained in Eq.~\ref{eq:scatt-rate} have been broadened using a Lorentzian linewidth with the lifetimes coming from the quasiparticles involved in the transition. We use Matthiessen's rule for computing the total broadening, that is, the broadening of the absorption is calculated as the sum of the linewidths of the excitons and the phonons involved in the scattering. The individual broadening have been taken from theoretical and experimental values in the literature\cite{kaasbjerg_unraveling_2012,park_electron-phonon_2008,spataru_theory_2005}. When not available, we fit them to the experimental absorption spectrum.
The values used in this work are shown in Table~\ref{tab:lifetimes}.
\begin{table}
\centering
\begin{tabular}{cccc|c}
\multicolumn{4}{c|}{Broadening (meV)} & \\ 
\hline
45.7 & 63.2 & 48.3  &      & $E_{11}$, $E_{22}$, $E_{12}$ \\
1.8 & 1.8 & 1.8 & 1.8 & $\mathbb{A}^3$, $\mathbb{A}^4$, $\mathbb{A}^5$, $\mathbb{A}^6$ \\
2.9 & 2.9 & 2.9 & 2.9 & $\mathbb{E}_1^3$, $\mathbb{E}_1^4$, $\mathbb{E}_1^5$, $\mathbb{E}_1^6$ \\
2.9 & 2.9 & 2.9 & 2.9 & $\mathbb{E}_2^3$, $\mathbb{E}_2^4$, $\mathbb{E}_2^5$, $\mathbb{E}_2^6$
\end{tabular}
\caption{Exciton and phonon broadenings.}
\label{tab:lifetimes}
\end{table}

Finally, we highlight that we do not calculate the scattering matrix elements, which are required as input in the Boltzmann equation. In this work we fit them to the amplitudes obtained from the experimental absorption spectrum (by minimising the difference with a Nelder-Mead algorithm).

\begin{figure} [tb]
    \centering
    \includegraphics[]{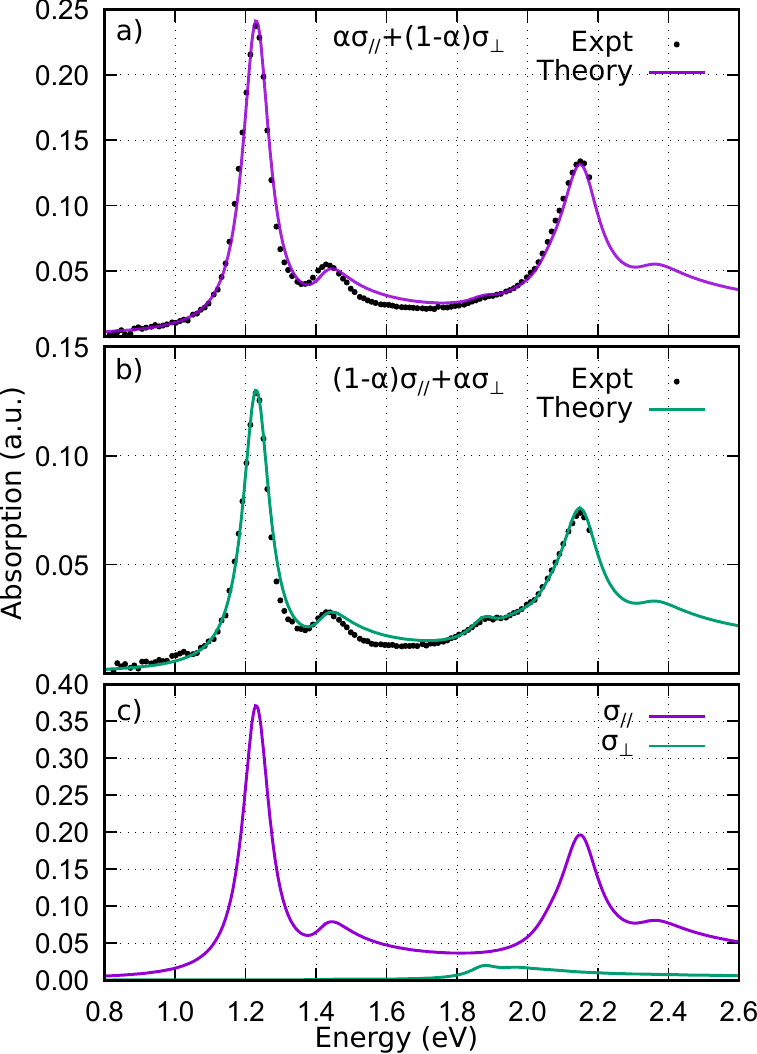}
    \caption{Total absorption spectrum of aligned (6,5) CNTs. Dotted curves are the experimental results. Solid lines are the fit obtained with our method. Figure a) and b) show the absorption spectrum of the CNT sample with parallel and perpendicular polarised light, respectively. Figure c) shows the individual contributions of the parallel and perpendicular absorption. The mixing parameter $\alpha=0.65$, corresponding to a nematic order parameter $S=0.3$.}
    \label{fig:total_abs}
\end{figure}

\textbf{Total absorption spectrum.}
The sample consists of single chirality aligned (6,5) SWCNTs. Absorption spectra for different polarisation angles have been measured (see Methods). In our work, we focus only on the parallel ($\beta=0$) and perpendicular ($\beta=\pi/2$) polarisation measurements.
Figure~\ref{fig:total_abs}a) and b) show the fit obtained with equations \ref{eq:total_para} and \ref{eq:total_perp}, respectively.
To estimate the degree of purity of the SWCNTs sample we decompose the absorption in its purely parallel and purely perpendicular components. These are shown in Figure~\ref{fig:total_abs}c).
From our fit, we obtain a mixing parameter $\alpha=0.65$, or equivalently a nematic order parameter of $S=0.3$.
The predominant $E_{11}$ and $E_{22}$ peaks are extremely well reproduced and exhibit the well known Lorentzian linewidth predicted by the theory. On the other hand, both the $E_{11}$ phonon side bands appearing in the perpendicular and parallel configuration are only partially reproduced by our approximation. Specifically, our description deviates slightly from the experimental measurements
in the region between 1.3 and 1.4 eV. The predicted side band peak position appears at higher energies than the experimental one. This is caused by our choice of using the \textit{ab initio} value of phonon energies provided in Table~\ref{tab:phonon}, which might differ in the case of a bundle of CNTs. 
Also, at energies above 1.4~eV, the theoretical curve overestimates the experimental absorption, which appears to decrease at a faster rate. This is most likely caused by the approximation of considering constant matrix elements and constant broadening for the scattering events.

Similarly to the $E_{11}$ peak, we predict the $E_{22}$ phonon sideband at around 2.3 eV in both Figure~\ref{fig:total_abs}a) and b), although we cannot compare with experimental data.
Finally, the cross polarisation peak due to the $E_{12}$ exciton is correctly reproduced at an energy around 1.9 eV. This is mostly visible in Figure~\ref{fig:total_abs}b) but gives a small contribution in a) as well. This is caused by the presence of misaligned SWCNTs in the sample. 

Lastly, we also note an important feature regarding the shallow absorption spectrum between 1.5 and 2.0~eV of Figure~\ref{fig:total_abs}a) and in the 1.5--1.8~eV region of Figure~\ref{fig:total_abs}b). The spectrum in this energy range emerges from the joint density of states of the phonon assisted $E_{11}$ scattering in Eq.~\ref{eq:scatt-rate}. We are able to reproduce the correct spectrum without using any polynomial baseline to fit the data. The latter approach is commonly used in experimental works where both exciton and exciton-phonon transitions are modelled as pure Lorentzians (or sometimes as Voigt lineshape) plus an unphysical polynomial background~\cite{pfohl_fitting_2017, katsutani_direct_2019}. Within our model, background subtraction methods are not needed. This will be more clear in the next section, where we will decompose the spectrum into its individual scattering contributions.

\begin{figure}[tb]
    \centering
    \includegraphics[]{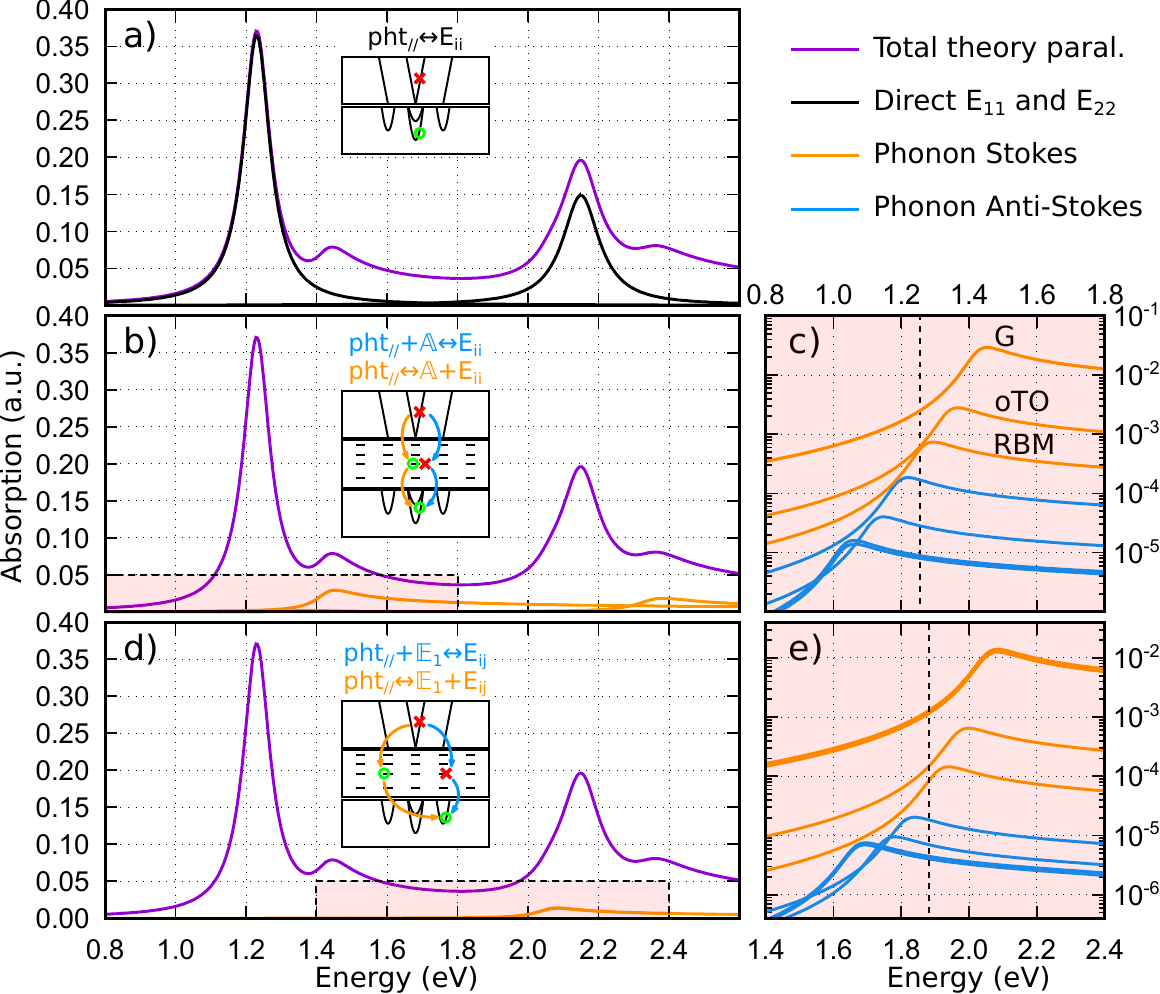}
    \caption{Purely parallel absorption spectrum. Left hand side plots show the different components of the spectrum as per Eq.~\ref{eq:parallel_scatt}. Right hand side plots are a magnification in logarithmic scale of the shade red regions in the main plots and show the contribution of the different phonon modes as of Table~\ref{tab:phonon}. Vertical dashed lines are the energy position of the exciton band minimum.}
    \label{fig:parallel}
\end{figure}

\begin{figure}[tb]
    \centering
    \includegraphics[]{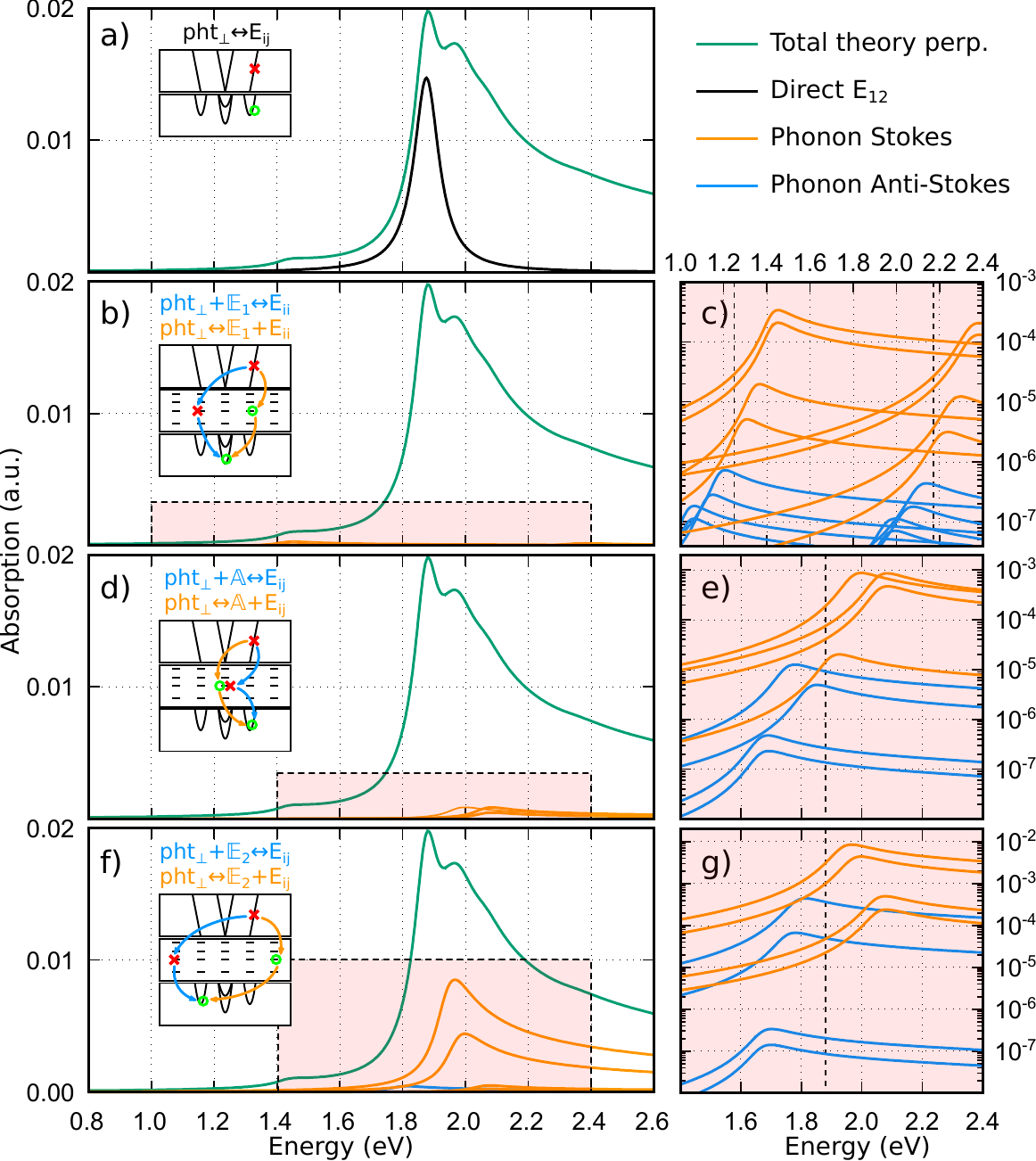}
    \caption{Purely perpendicular absorption spectrum. Left hand side plots show the different components of the spectrum as per Eq.~\ref{eq:perpendicular_scatt}. Right hand side plots are a magnification in logarithmic scale of the shade red regions in the main plots. Vertical dashed lines are the energy position of the exciton band minimum.}
    \label{fig:perpendicular}
\end{figure}

\textbf{Stokes and anti-Stokes processes.}
The left-hand side of Figs.~\ref{fig:parallel} and \ref{fig:perpendicular} show the decomposition of the purely parallel and purely perpendicular absorption spectra into the scattering families of equations \ref{eq:parallel_scatt} and \ref{eq:perpendicular_scatt}.
The top left panel in both figures displays the contribution to the spectrum due to the direct absorption of photons. The panels below show the contributions coming from phonons at different regions of the helical BZ. Each panel has an inset with a small pictorial representation of the corresponding transition. In these insets, some of the exciton, phonon or photon bands are connected by coloured arrows showing one of the possible excitation paths. Red crosses and green circles represent the annihilation or the creation of a particle, respectively. Phonon assisted scatterings are divided in two categories, Stokes and anti-Stokes processes, which represent the emission or the absorption of a phonon, respectively. The right hand side sub-figures magnify the shaded red area of the main plots. These sub-figures are plotted on logarithmic scale and have a vertical dashed line indicating the energy of the bottom of the corresponding exciton band.

Figure~\ref{fig:parallel}a) shows the part of the absorption spectrum of parallel polarised light due to two transitions: the direct excitation of $E_{11}$ and $E_{22}$ exciton. Interestingly, while the $E_{11}$ peak accounts for almost all of the total absorption spectrum at 1.23~eV, the $E_{22}$ exciton is only responsible for roughly 3/4 of the total absorption at 2.15~eV, since a large tail of lower energy phonon side bands reaches these energies. Figure~\ref{fig:parallel}b) shows the contribution to absorption due to generation of $E_{11}$ and $E_{22}$ assisted by $\mathbb{A}$ phonons. Figure~\ref{fig:parallel}d) instead describes the excitation of the cross-polarised exciton $E_{12}$ mediated by the $\mathbb{E}_1$ phonons. Differently from direct transitions, the lineshape of these contributions are not Lorentzian: the peak position differs from the exciton by the energy of the phonons and, most importantly, have a long asymmetric tail that stretches to higher energies. These tails arise from the joined density of states of the transition involving non-zero phonon momentum transfer to higher-energy excitonic states.

Subplots \ref{fig:parallel}c) and \ref{fig:parallel}e) show in logarithmic scale the individual contributions. We see that $\mathbb{A}$ phonon-assisted scatterings have the same order of magnitude of $\mathbb{E}_1$ ones. We can also resolve the individual contributions within a given channel, however, let us warn that uncertainties on the phonon energies in the real sample, make the decomposition into individual contributions less reliable. G+ and G-- phonons contribute the most to the total sideband absorption. As expected, anti-Stokes scattering at room temperature play only a negligible role in the total absorption but are included as they will become important in excited samples with out-of-equilibrium phononic population. Stokes and anti-Stokes peaks are positioned symmetrically with respect to the exciton energies, their lineshape is not Lorentzian and their magnitude depends directly on the phonon population~\cite{souza_filho_electronic_2001,yu_phonon_2010}. 

We now turn to the perpendicular direction. Figure~\ref{fig:perpendicular} a) shows the direct excitation of the cross polarised $E_{12}$ exciton due to perpendicular polarised light. Similarly to the parallel case, it has a Lorentzian shape with 48.3~meV and accounts for almost the entire absorption at around 1.9~eV.
The magnitude of the $E_{12}$ peak is small compared to $E_{11}$ due to the depolarisation effect pointed out by Ajiki and Ando~\cite{ajiki_aharonov-bohm_1994,ando_theory_2005}. In our work, the ratio between $E_{12}$ and $E_{11}$ amplitudes is 0.043. Phonon assisted transitions are shown in subfigures \ref{fig:perpendicular}b), d) and f). In the first subplot we observe a small but non negligible contribution of $\mathbb{E}_1$ phonons to the two parallel polarised excitons, which causes a small shoulder at 1.4~eV to appear. The last two sublplot are similar in behaviour and show the phonon sidebands of the $E_{12}$. $\mathbb{A}$ phonons are one order of magnitude smaller than $\mathbb{E}_2$ phonons, the lineshape is not Lorentzian with a long tail towards higher energies.
Finally, we observe that oTO, G+ and G-- $\mathbb{A}$ and $\mathbb{E}_2$ symmetry phonons are the ones contributing more to the absorption compared to $\mathbb{E}_1$.


\section{Conclusions}
We developed a model for the optical absorption of SWCNTs based on the Boltzmann scattering equation. 
No assumptions are made on the particle populations, and the scattering integrals are calculated exactly using a recently developed algorithm. We were able to reproduce our experimental optical spectra of aligned (6,5) SWCNTs with a high degree of confidence and identify each single scattering contribution. Notably, no background baseline was required in order to reproduce the experimental data.  We attribute this baseline to phonon-assisted continuum absorption due to transitions to finite-momentum excitonic states.
This first critical step, obtained with an approach and a numerical method that can handle strongly out-of-equilibrium situations, indicates that, in our future work, time-dependent spectra can be constructed and provide a very accurate insight into the dynamics of not only the electronic system in SWCNTs, but also the transient population evolution of the phonon subsystem.

\section{Methods}

\textbf{Purification.} (6,5) SWCNTs (Signis\textregistered SG65i, Sigma-Aldrich) were separated by a gel-chromatography method~\cite{yomogida_industrial-scale_2016}. After suspending SWCNTs in an aqueous solution with sodium cholate (SC, Sigma-Aldrich) and sodium dodecyl sulfate (SDS, Sigma-Aldrich) by tip sonication followed by ultracentrifugation, gel-chromatography was used to first separate semiconducting SWCNTs by chiral angles under the surfactant environment of 2.0\% (wt./vol.) SDS and 0.5\% SC, and then to sort them by diameters by changing the concentration of sodium deoxycholate (DOC, Sigma-Aldrich) in 0.5\% SDS and 0.5\% SC solution.\cite{ichinose_extraction_2017}
The purity of (6,5) SWCNTs obtained by this process was ~75\%, and this solution was used to produce aligned films.

\textbf{Aligned Film Fabrication.}
Aligned films of SWCNTs were produced by a vacuum filtration method~\cite{komatsu_groove-assisted_2020, he_wafer-scale_2016}. Purified (6,5) suspension was re-dispersed in 0.04\% DOC solution. After further diluting the solution with pure water (typical DOC concentration $\sim$ 0.01\%), we filtered the solution using the filtration set up (Microanalysis Filter Holder, Millipore) with 80 nm pore size filter membranes (Nuclepore Track-Etched Polycarbonate (Hydrophilic) Membranes, GE Healthcare Life Sciences) under a well-controlled filtration speed. Films were transferred onto glass substrates by a wet transfer method, where the filter membranes were dissolved by chloroform (Thermo Fisher Scientific) and then rinsed by 2-propanol (Thermo Fisher Scientific) and pure water.

\textbf{Absorption Measurements.}
The attenuation spectra of CNT films were measured by a home-built optical measurement setup, consisting of a tungsten-halogen lamp (SLS201L, Thorlabs), a Glan-Thompson polarizer, and two spectrometers (covering 567-1050 nm and 1050-1568 nm, respectively), similar to Ref.~\cite{katsutani_direct_2019}. The former one consists of a monochromator (Horiba/JY, Triax320) equipped with a liquid-nitrogen-cooled CCD camera (Princeton Instruments, Spec-10), and the latter one consists of a monochromator (Princeton Instruments, SP-2150) with a liquid-nitrogen-cooled 1D InGaAs detector array (Princeton Instruments, OMAV InGaAs System). The beam size was ~1 mm. We rotated the polarizer to change the angle between the incident light and the CNT alignment direction. The attenuation coefficient is calculated as
\begin{equation}
A = -\log_{10}\left( \frac{T_{sample}}{T_{ref}} \right),
\end{equation}
where $T_{sample}$ and $T_{ref}$ are the transmission coefficient of the CNT and that of the substrate, respectively.

\textbf{Quantum Fokker-Planck equation.}
A generic collision integral in the Boltzmann equation is written in terms of the quantum Fokker--Plank equation \cite{snoke_solid_2020}.
For instance, in the case of two particles in an initial state $|12 \rangle$ with wavevector $\mathbf{k}_1$ and  $\mathbf{k}_2$ and populations $f_1$ and $ f_2$ that scatter into a final state $|34 \rangle$ with wavevector $\mathbf{k}_3$ and $\mathbf{k}_4$ and populations $f_3$ and $ f_4$, the rate of change of population $f_1$ as a consequence of the interaction with $f_2$, $f_3$ and $f_4$ is described by the following scattering collision integral
\begin{equation}
    {\left( \frac{ \partial f_1}{ \partial t} \right)}_{\text{12--34}} = \frac{2\pi}{\hbar} \frac{1}{V^2_{BZ}} \sum_{\mathbf{G}}  \iiint_{V^3_{BZ}} d\mathbf{k}_2 d\mathbf{k}_3 d\mathbf{k}_4 \; |g_{12}^{34}|^2 \; \,\delta_{\mathbf{k}}\; \delta_{\epsilon}\; P_{12}^{34},
    \label{eq:quantum-Fokker-Plank}
\end{equation}
where $g_{12}^{34} = \langle 12| \hat{V}| 34 \rangle = g(\mathbf{k}_1, \mathbf{k}_2, \mathbf{k}_3, \mathbf{k}_4 )$ are the matrix elements of the interaction $\hat{V}$ between the initial and the final state.
Energy and momentum conservation are enforced by
\begin{equation}
\begin{split}
    \delta_{\mathbf{k}} &= \delta (\mathbf{k}_1 + \mathbf{k}_2 - \mathbf{k}_3 - \mathbf{k}_4 + \mathbf{G}), \\
      \delta_{\epsilon} &= \delta (\epsilon_1(\mathbf{k}_1) + \epsilon_2(\mathbf{k}_2) - \epsilon_3(\mathbf{k}_3) - \epsilon_4(\mathbf{k}_4)),
    \label{eq:dirac-deltas}
\end{split}
\end{equation}
where $\epsilon_i(\mathbf{k}_i)$ are the energy dispersion relations of the (quasi)particles considered, and the sum over the reciprocal lattice vectors $\mathbf{G}$ accounts for Umklapp scatterings.
Finally the term $P_{12}^{34}$ accounts for the particle populations as
\begin{equation}
    P_{12}^{34} = (1 \pm f_1) (1 \pm f_2) f_3 f_4 - f_1 f_2 (1 \pm f_3) (1 \pm f_4),
\end{equation}
where the plus and minus signs apply to bosonic and fermionic particles respectively, while the first and second term represent the direct and time-reversed process respectively. The expression $f_i$ stands for $f(\mathbf{k}_i)$.

Starting from the quantum Fokker-Plank equation Eq.~\ref{eq:quantum-Fokker-Plank}, we can construct the scattering rate, or inverse lifetime, for any of the particles involved in the transition. This is achieved by performing a functional derivative with respect to $f_1$ at equilibrium. Here we simply state the final result, that reads
\begin{equation}
    \frac{1}{\tau(\mathbf{k}_1)} = \frac{\delta}{\delta f_1} {\left( \frac{ \partial f_1}{ \partial t} \right)}_{\text{12--34}} = \frac{2\pi}{\hbar} \frac{1}{V^2_{BZ}} \frac{1}{2} \sum_{\mathbf{G}} \int \int \int_{V^3_{BZ}} d\mathbf{k}_2 d\mathbf{k}_3 d\mathbf{k}_4 |g_{12}^{34}|^2 \delta_{\mathbf{k}} \delta_{\epsilon} P.
\end{equation}
The factor $P$ is
\begin{equation}
    P = (1 \pm f_2) f_3 f_4 + f_2 (1 \pm f_3) (1 \pm f_4),
\end{equation}
where the plus or minus indicate bosons or fermions, respectively.
For a more extensive introduction and for details about the numerical implementation see Refs. \cite{wais_numerical_2021,wadgaonkar_numerical_2021-1,wadgaonkar_numerical_2021,bagsican_terahertz_2020,wais_quantum_2018} or the Supplementary Information. 

In this work we model the absorption of light using this framework. The rate of change of the photon population is given by the scattering terms of the BE. This should not be surprising, since the quantum Fokker--Plank equation is nothing else than the generalisation of Fermi's golden rule, which is routinely used for light-matter interaction\cite{dresselhaus_exciton_2007}.

\begin{suppinfo}
Supporting information of CNT absorption.
\end{suppinfo}

\begin{acknowledgement}
S.D.F., M.W., I.W. and M.B. acknowledge the Nanyang Technological University, Singapore, grant NAP SUG.
N.K., A.M., S.G., and J.K. acknowledge support by the Basic Energy Science (BES) program of the U.S. Department of Energy through Grant No. DE-FG02-06ER46308 (for preparation of aligned carbon nanotube films) and the Robert A. Welch Foundation through Grant No.\ C-1509 (for structural characterisation measurements).
\end{acknowledgement}

\bibliography{CNTabs}

\end{document}